\documentclass[twocolumn]{webofc}
\usepackage[varg]{txfonts}
\usepackage{hyperref}
\usepackage{url}
\usepackage{listings}
\hypersetup{colorlinks=true,citecolor=blue,urlcolor=blue,linkcolor=blue}
\newcommand{\EBz}{Z_\text{B}}
\newcommand{\Rin}{R_\text{inner}}
\newcommand{\PHISEGMENTS}{N_\Phi}
\newcommand{\nomfw}{C_\text{fw}}
\newcommand{\nomth}{T_\text{th}}
\newcommand{\Fdz}{F_\text{dz}}
\newcommand{\Rdz}{R_\text{dz}}
\newcommand{\sipmth}{S_\text{th}}

\newcommand{\DPHIGLOBAL}{d\Phi}
\newcommand{\THETASIZEBARREL}{\Theta_B}
\newcommand{\THETASIZEENDCAP}{\Theta_E}
\newcommand{\NTHETABARREL}{N\theta_B}
\newcommand{\NTHETAENDCAP}{N\theta_E}
\newcommand{\DTHETABARREL}{d\theta_B}
\newcommand{\DTHETAENDCAP}{d\theta_E}
\newcommand{\NPHIBARRELCRYSTAL}{N\phi_B}
\newcommand{\NPHIENDCAPCRYSTAL}{N\phi_E^*}
\newcommand{\DPHIBARRELCRYSTAL}{d\phi_B}
\newcommand{\DPHIENDCAPCRYSTAL}{d\phi_E^*}

\begin{document}
\title{Differentiable Full Detector Simulation of a Projective Dual-Readout Crystal Electromagnetic Calorimeter with Longitudinal Segmentation and Precision Timing}
\author{\firstname{Wonyong} \lastname{Chung}\inst{1}\thanks{\email{wonyongc@princeton.edu}}}
\institute{Department of Physics, Princeton University, New Jersey, USA}
\abstract{
A differentiable full detector simulation has been implemented in the key4hep software stack for future colliders. A fully automated and configurable geometry enabling differentiation of all detector dimensions, including crystal widths and thicknesses, is presented. The software architecture, development environment, and necessary components to implement a new detector concept from scratch are described. General AI/ML reconstruction strategies for future collider detectors are discussed, based around the idea of picking the right neural network for each detector.
}
\maketitle
\section{Differentiable Simulation}
\label{intro}
So-called differentiable simulation is a recent and active area of research which aims to develop dynamically re-configurable detector geometries to be used in conjunction with analysis templates in a feedback loop against physics benchmarks. Such a setup would allow for bilevel optimization of reconstruction hyperparameters and granular detector dimensions, such as crystal sizes, which were once difficult to study systematically and iteratively in a full detector geometry. Here we present the first such differentiable full detector simulation for projective geometries.

This work specifically addresses the technical implementation of such a detector in the latest key4hep software infrastructure~\cite{key4hep}, which unifies the software dependencies and simulation programs for all future collider experiments. The use of so-called compact XML files in the detector description toolkit for high-energy physics (dd4hep) to define the high-level placements of subdetectors already allows for a high level of configurability in re-arranging and interchanging subdetectors. Certain parameters, such as the number of layers in a tile-based sampling calorimeter, may also be specified in the compact XML file, allowing for dynamic re-configuration, provided the detector construction routine has implemented the detector elements in question in a re-configurable manner. Material properties may also be changed in the compact XML.

In calorimetry, dd4hep currently provides cylindrical geometries in the nomenclature of staves and layers, used in certain tracker designs and tile-based sampling calorimeters. To implement a new detector concept based on instead a projective geometry, which is not currently implemented in dd4hep, it is necessary to write the projective geometry calculations from scratch to construct the individual detector element volumes and rotate and place them in the correct positions with exact hermiticity.

Given the absence of projective geometries in dd4hep, we take the opportunity to write a fully automated and dynamically configurable projective geometry with longitudinal segmentation capabilities, wherein all detector dimensions are dynamically calculated from only a few input specifications. This, combined with the requisite sensitive action and hit class definitions customized for dual-readout, completes the implementation of the first fully differentiable calorimeter design for future colliders.

\begin{figure}[h]
\centering
\includegraphics[width=8cm,clip]{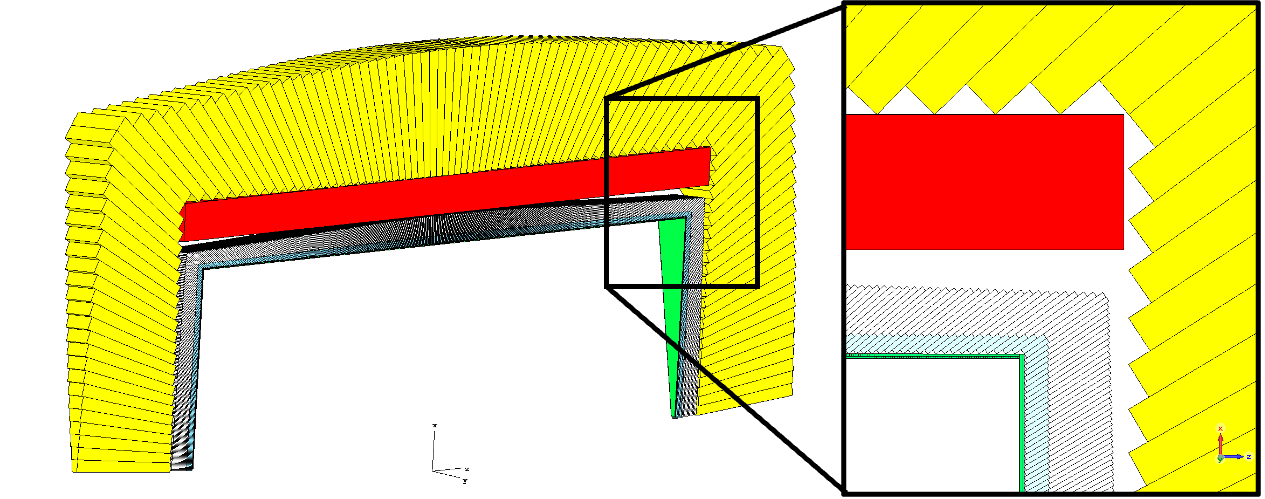}
\caption{
A single phi segment of a combined calorimeter concept with a precision timing layer (green) and segmented crystal layer (front crystals: blue, rear crystals: white) comprising the ECAL placed in front of the solenoid (red) of an IDEA-like dual-readout fiber HCAL (yellow towers; fibers not shown). (Insert) Cross-sectional view of the barrel/endcap junction showing the ECAL, solenoid, and HCAL towers.
}
\label{phisegment}
\end{figure}

\begin{table*}[t]
\centering
\caption{Input parameters for parameterized geometry construction.}
\label{input_parameters}
\begin{tabular}{lll}
\hline
Description                                         & Variable Name      & Value  \\\hline
Half Z-extent of the barrel                         & $\EBz$             & 2.25\,m \\
Inner radius of the barrel                          & $\Rin$             & 2\,m \\
Global number of phi segments                       & $\PHISEGMENTS$     & 64 \\
Nominal square face width of the front crystals     & $\nomfw$           & 1\,cm \\
Nominal square thickness of timing crystals         & $\nomth$           & 3\,mm \\
Front crystal length                                & $\Fdz$             & 5\,cm \\
Rear crystal length                                 & $\Rdz$             & 15\,cm \\
SiPM wafer thickness                                & $\sipmth$          & 0.5\,mm \\\hline
\end{tabular}
\end{table*}

\begin{table*}[t]
\centering
\caption{Secondary parameters calculated from input parameters.}
\label{secondary_parameters}
\begin{tabular}{lll}
\hline
Description                                                   & Variable Name         & Formula  \\\hline
Angular size of a single phi segment                          & $\DPHIGLOBAL$         & 2$\pi$/ $\PHISEGMENTS$ \\
Angular size of barrel region                                 & $\THETASIZEBARREL$    & atan($\EBz/\Rin$) \\
Angular size of endcap region                                 & $\THETASIZEENDCAP$    & atan($\Rin/\EBz$) \\
Number of barrel segments in $\theta$                         & $\NTHETABARREL$       & floor($2\EBz/\nomfw$) \\
Number of endcap segments in $\theta$                         & $\NTHETAENDCAP$       & floor($\Rin/\nomfw$) \\
Angular size of a single barrel segment in $\theta$           & $\DTHETABARREL$       & $(\pi-2\THETASIZEENDCAP)/\NTHETABARREL$ \\
Angular size of a single endcap segment in $\theta$           & $\DTHETAENDCAP$       & $\THETASIZEENDCAP/\NTHETAENDCAP$ \\
Number of barrel segments in $\phi$ in a single phi segment   & $\NPHIBARRELCRYSTAL$  & floor($2\pi\Rin/(\PHISEGMENTS\nomfw)$) \\
Number of endcap segments in $\phi$ in a single phi segment$^*$ & $\NPHIENDCAPCRYSTAL$  & floor($2\pi\Rin^*/(\PHISEGMENTS\nomfw)$) \\
Angular size of barrel segments in $\phi$                     & $\DPHIBARRELCRYSTAL$  & $\DPHIGLOBAL/\NPHIBARRELCRYSTAL$ \\
Angular size of endcap segments in $\phi$                     & $\DPHIENDCAPCRYSTAL$  & $\DPHIGLOBAL/\NPHIENDCAPCRYSTAL$ \\\hline
\end{tabular}
\end{table*}

\section{Segmented Crystal EM Precision Calorimeter (SCEPCal)}
Dual-readout calorimetry has a recent but rich history~\cite{25years} and offers a genuinely new type of calorimetry for use in future collider detectors. First pioneered by the RD52 experiment and proposed for a future lepton collider by the IDEA collaboration~\cite{idea}, the benefits of improving EM resolution by adding a longitudinally-segmented homogeneous crystal electromagnetic calorimeter in front of the IDEA detector has been studied in several recent works~\cite{Lucchini_2020,calorimetryatfccee}.

\begin{figure}[h]
\centering
\includegraphics[width=8cm,clip]{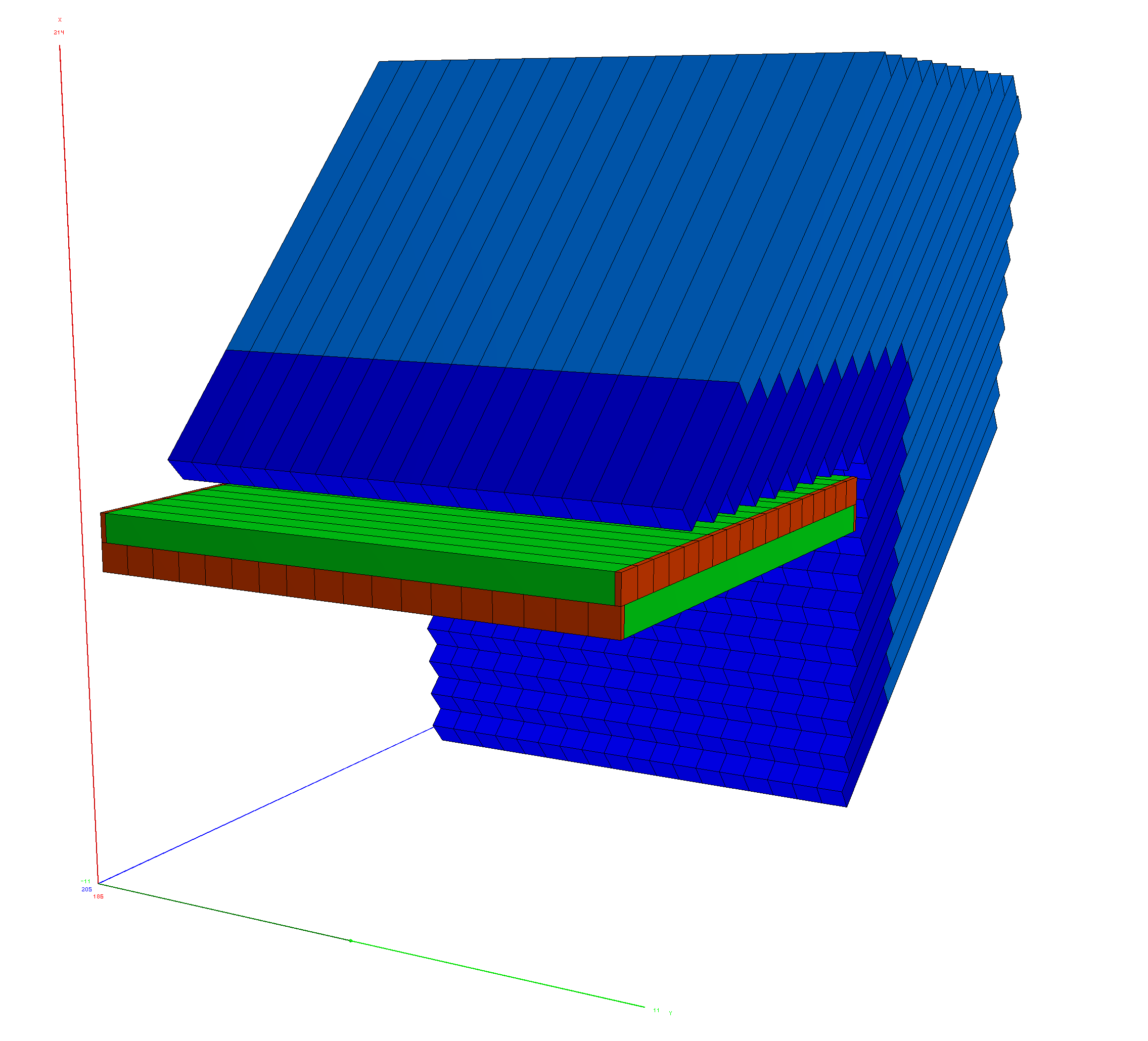}
\caption{
Barrel/endcap junction of a single phi segment. The projective layer (front crystals: dark blue, rear crystals: light blue) is further segmented into $\NPHIBARRELCRYSTAL$ and $\NPHIENDCAPCRYSTAL$ segments according to the nominal face width $\nomfw$. A single timing layer tile is shown, with crystals (green) arranged in an alternating fashion. SiPM wafers (red) are attached to both ends of each timing crystal. Dimensions exaggerated for visibility.
}
\label{gammasegments}
\end{figure}

The SCEPCal~\cite{Lucchini_2020} is a projective geometry of homogeneous crystals much like the CMS or L3 detectors, but with each crystal longitudinally segmented into a front and rear compartment, with the lengths in a ratio of 6:16 radiation lengths, totaling 22\,$X_0$ for full EM shower containment. SiPMs are instrumented to the front and rear faces of the respective segments to avoid the introduction of dead material between the crystals, with separate SiPMs for scintillation and Cerenkov light instrumented only on the rear crystals since hadrons tend to shower late in the crystal. A precision timing layer is placed in front of the projective crystal layer, comprised of two layers of thin, fast-scintillating crystals arranged in square tiles of alternating orientation, with SiPMs instrumented at each end of the crystals similar to the CMS BTL~\cite{cmsmtdtdr}. The layers are shown in the insert of Figure~\ref{phisegment}, which shows a single phi segment of the detector.

Lead tungstate (PbWO$_\text{4}$) is used for the projective layer crystals, and LYSO for the timing layer. LYSO is used as a baseline because it is used in the CMS BTL, but is likely to change as its performance in a lepton collider environment is studied. Material properties have a significant impact on simulation results and remain an active area of research.

\section{Projective Geometry}
A idealized projective geometry can be fully determined and parameterized with the parameters listed in Table~\ref{input_parameters}, which are then used to calculate the secondary parameters listed in Table~\ref{secondary_parameters}. The inner radius, Z-extent of the barrel, and number of total segments in phi determine most of the detector dimensions, and the nominal values of the crystal face-widths and thicknesses are used to subdivide larger detector dimensions into integer numbers of projective crystals in an exactly hermetic manner with no overlaps. The side lengths and orientations for each asymmetric trapezoidal volume in the projective geometry are then calculated automated and placed in their respective positions.

\subsection{Projective Gaps}
In this configuration, the center of the inner surface of a global phi segment is oriented normal to the interaction point (IP), and the further segmentation in phi introduces an offset distance from the center while leaving the face orientation unchanged, thus accounting for projective gaps in all but the centermost crystals of a global phi segment. The projective gaps for these crystals are remedied by introducing additional $xy$-slices of non-projective, constant-angle rings of crystal at $z=0$ to push out the projection from the IP.

\begin{figure}[h]
\centering
\includegraphics[width=8cm,clip]{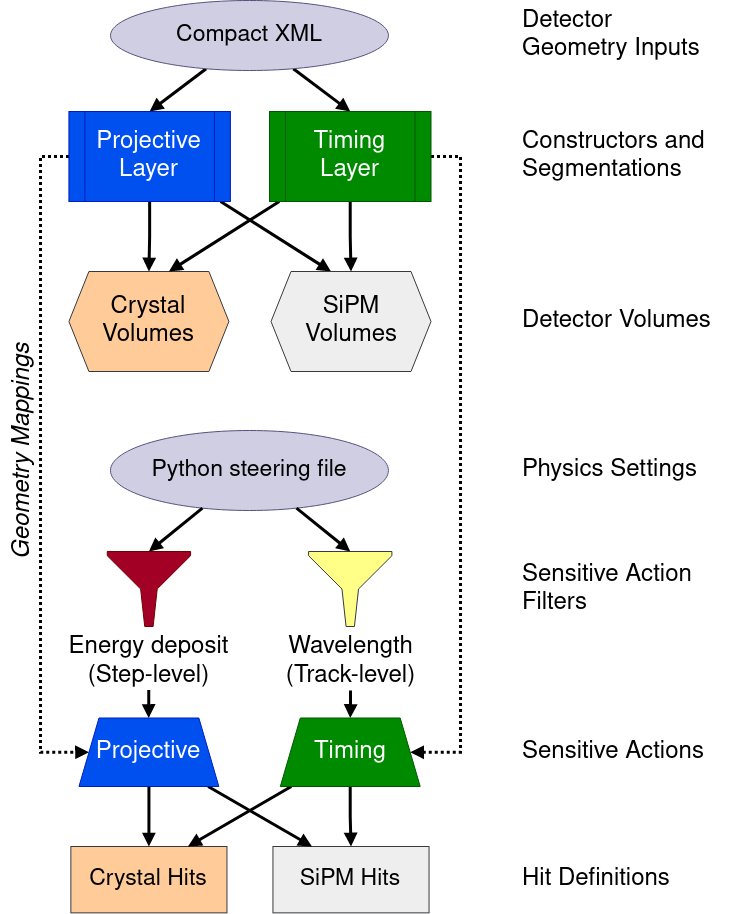}
\caption{
Schematic diagram of software components for a full detector simulation in dd4hep.
}
\label{software}
\end{figure}

\section{Full Simulation}
The necessary software components for a new detector concept implementation in dd4hep are shown schematically in Figure~\ref{software}. The compact XML file specifies the subdetectors to be used and their parameters, as well as material properties and visualization attributes. The construction routines for each subdetector calculate, generate, and place the physical volume elements in Geant4. Utility classes known as segmentations provide ancillary information such as global position coordinates for detector cells when writing out hits. A python steering file sets physics settings not directly related to the geometry, such as the physics process list, selection cuts, and event generation settings. The steering file also includes simulation-related settings such as logging, random number seeds, and threshold cuts for sensitive action filters, which determine the particles for which information is saved in the subdetector data writeout. Custom sensitive action classes and hits must be written for a new detector concept if existing ones, which record only positions and energy deposits, are not sufficient to capture information generated by the new detector, such as for SiPMs in dual-readout calorimeters, which record the digitized wavelength and time bins of incident scintillation and Cerenkov photons.

\subsection{Development Environment}
Visualization is a critical aspect of new detector design development. While technically several options exist, latencies at the order of 100\,ms or less are desired for the minimally adequate experience. For maximum compatibility and responsiveness we recommend a local CERN-like setup: AlmaLinux 9 with HEP\_OSlibs~\cite{heposlibs} and key4hep-spack~\cite{key4hep}.

\section{Reconstruction}

Detectors at future colliders are widely expected to run real-time inference on front-end ASICs for so-called triggerless operation, and development of AI/ML reconstruction algorithms based on the latest neural network architectures is a fast-emerging area with wide interest~\cite{gravnet,mlpf}. In recent years the trend has been towards treating clusterization as an inner detail of overall reconstruction of unordered particle clouds~\cite{particlenet}, in contrast to the delineated approach towards jets used in rule-based Particle Flow algorithms (PFAs)~\cite{cmspfa}. 

A differentiable full simulation opens doors to a multitude of new types of studies in which granular components of the detector design can be optimized systematically against selected physics metrics. Optimizing the geometry against the reconstruction algorithms, which themselves are simultaneously being optimized, is surely a task for AI/ML. Even still, neural network architectures themselves are also the subject of intense research and are usually conceived and developed to solve specific classes of problems intuitively representing use-cases such as image recognition (convolutional neural networks) and natural language processing (transformers). In collider physics, recent work has tended toward graph neural networks for their ability to represent relationships between detector objects like calorimeter hits to physics objects and their kinematics. 

However, networks trained solely on static or parameterized fast-simulation models are inherently inefficient as insight into intrinsic detector characteristics that cannot be parameterized, like angular resolution, material properties, and timing information, is lost, thus precluding bilevel optimization of reconstruction hyperparameters against geometry parameters.

\subsection{Picking the Right Neural Network for the Detector}
Particularly in calorimetry, the physical characteristics of a detector must be taken into account when considering the type of neural network architecture best suited for reconstruction of its readouts. Different detectors have different strengths, and hits from a silicon-tungsten sandwich calorimeter cannot be considered in the same, monolithic class of objects as hits from a homogeneous, but slower and coarser, crystal apparatus, for example. The original class of problems for which a network architecture was designed should also be taken into account in view of its compatibility to the class of physics questions which are at hand in reconstruction.

A salient example is in the case of diffusion models~\cite{diffusion}, which inherently reconstruct from noise and differ fundamentally from the lineage of vector-token architectures like the transformer~\cite{transformers}. Recent extensions to diffusion models like ControlNet~\cite{controlnet}, which allows for spatial contextual-control of model outputs, evoke strong parallels to reconstruction tasks in pixelated sampling calorimeters, which have excellent mechanical resolution but high noise. Such an implementation would use tracks as the spatial contextual-control against the stochastic structure of the noisy calorimeter hits, which the diffusion model has learned.

For calorimeters with granular longitudinal separation capabilities between EM showers and jets, such as with dual-readout crystals and fiber towers, additional handles such as precision timing information contribute to a long-range sequence of data suitable for use in a transformer architecture. Using tracks in the attention-step could potentially alleviate the short-range contextual limits that traditional graph neural networks are known to suffer from.

Different network architectures may also be combined within a single detector when applicable, such as in concepts like the Crilin detector, which features both a sampling and homogeneous portion. A diffusion model could be used in the sampling portion, and a transformer model in the homogeneous section, and a generative adversarial network used as a controller to integrate the results of both networks into a final reconstructed object.

\bibliography{sn-bibliography}
\end{document}